\documentclass[a4paper,11pt]{article}
\usepackage[margin=2.5cm]{geometry}
\usepackage{setspace,graphicx,url,hyperref,enumitem}
\usepackage[linesnumbered,ruled]{algorithm2e}

\usepackage{amsmath,amssymb,amsfonts,amsthm,natbib}

\newcommand{\R}{\ensuremath{\mathbb{R}}}
\newcommand{\B}{\ensuremath{\mathbb{B}}}

\begin{document}
\onehalfspacing
\title{Boolean Hierarchical Tucker Networks on Quantum Annealers}
\author{Elijah Pelofske\footnote{Los Alamos National Laboratory, Los Alamos, NM 87545, USA}~, Georg Hahn\footnote{Harvard University, T.H.\ Chan School of Public Health, Boston, MA 02115}~, Daniel O'Malley\footnotemark[1]~,\\Hristo N.\ Djidjev\footnotemark[1]~, and Boian S.\ Alexandrov\footnotemark[1]}
\date{}
\maketitle

\begin{abstract}
    Quantum annealing is an emerging technology with the potential to solve some of the computational challenges that remain unresolved as we approach an era beyond Moore's Law. In this work, we investigate the capabilities of the quantum annealers of D-Wave Systems, Inc., for computing a certain type of Boolean tensor decomposition called Boolean Hierarchical Tucker Network (BHTN). Boolean tensor decomposition problems ask for finding a decomposition of a high-dimensional tensor with categorical, [true, false], values, as a product of smaller Boolean core tensors. As the BHTN decompositions are usually not exact, we aim to approximate an input high-dimensional tensor by a product of lower-dimensional tensors such that the difference between both is minimized in some norm. We show that BHTN can be calculated as a sequence of optimization problems suitable for the D-Wave 2000Q quantum annealer. Although current technology is still fairly restricted in the problems they can address, we show that a complex problem such as BHTN can be solved efficiently and accurately.
\end{abstract}

\section{Introduction}
\label{sec:intro}
One of the most powerful tools for extracting latent (hidden) features from data is factor analysis \citep{spearman1961general}. Traditionally, factor analysis approximates some data-matrix $X \in \R^{n \times m}$ by a product of two factor matrices, $X \approx AB$, where $A \in \R^{n \times k}$, $B \in \R^{k \times m}$, and $k \ll n,m$. Various factorizations can be obtained by imposing different constraints.

Assuming the input data is a matrix  $X \in \B^{n \times m}$, where $\B=\{0,1\}$, Boolean factorization is looking for two binary matrices $A \in \B^{n \times k}$, $B \in \B^{k \times m}$ such that $X = AB$, where $x_{ij}=\vee_{l=1}^k a_{il}y_{lj} \in \B$ and $\vee$ is the logical "or" operation ($1+1=1$). The \textit{Boolean rank} of $X$ is the smallest $k$ for which such an exact representation, $X=AB$, exists. Interestingly, in contrast to the non-negative rank, the Boolean rank can be not only bigger or equal, but also much smaller than the logarithm of the real rank \citep{Monson1995, desantis2020factorizations}. Hence, low-rank factorization for Boolean matrices is  of particular interest. Past works on utilizing quantum annealing for matrix factorization were focused mostly on non-negative matrix factorizations \cite{o2018nonnegative,golden2021reverse}.

Instead of using matrix representation, many contemporary datasets are high-dimensional and need to be represented as \textit{tensors}, that is, as multidimensional arrays. Tensor factorization is the high-dimensional generalization of matrix factorization \citep{kolda2009tensor}. In this work, we present an algorithm for Boolean Hierarchical Tucker Tensor Network computation using the D-Wave 2000Q annealer. \textit{Boolean Hierarchical Tucker Tensor Network (BHTN)} is a special type of tensor factorization, in which tensor product operations are structured in a tree-like fashion. We start by reshaping/unfolding a high-dimensional tensor into an equivalent matrix representation. Afterwards, this matrix representation is decomposed recursively by reshaping and splitting using Boolean matrix factorization. We show that the factorization problem of Boolean matrices, which is the most computationally challenging step, can be reformulated as a \textit{quadratic unconstrained binary optimization (QUBO)} problem, which is a type of problem suitable for solving on D-Wave's quantum annealer. Problems previously solved on D-Wave include maximum clique, minimum vertex cover, graph partitioning, and many other NP-hard optimization problems. The D-Wave 2000Q device situated at Los Alamos National Laboratory has over 2000 qubits, but because of the limited connections (couplers) between those qubits, it is usually limited to solving optimization problems of sizes (number of variables) not higher than 65. Nevertheless, due to the special structure of our algorithm, we are able to compute BHTNs of much larger tensors. The largest one we tried had $65536$ elements.

Our contribution is threefold: First, we present a recursive algorithm that calculates a BHTN. Second, we reformulate the Boolean matrix factorization as a higher-order binary optimization (HUBO) problem. Third, after having converted the HUBO to a QUBO, suitable for processing via quantum annealing, we use  D-Wave 2000Q to compute BHTNs for input Boolean tensors of various orders, ranks,  and sizes. No quantum algorithms for computing BHTNs have been previously published.

The article is structured as follows: Section~\ref{sec:methods} starts by presenting a high-level overview of Hierarchical Tucker Networks, which allow us to recursively decompose tensors into a series of lower-order tensors. We show that on each level of the recursion, we are faced with factorizations of Boolean matrices, a problem which is equivalent to minimizing a higher-order binary optimization problem that can be solved on D-Wave. We evaluate our algorithm on different input tensors in Section~\ref{sec:experiments}. The article concludes with a discussion in Section~\ref{sec:discussion}.

\section{Methods}
\label{sec:methods}
This section describes all components of the Boolean Hierarchical Tucker Network leading to a higher-order binary optimization problem, and the subsequent solving step on D-Wave.

\subsection{Hierarchical Tucker decomposition}
\label{sec:tucker}
\begin{figure}[t]
    \centering
    \includegraphics[width=0.7\textwidth]{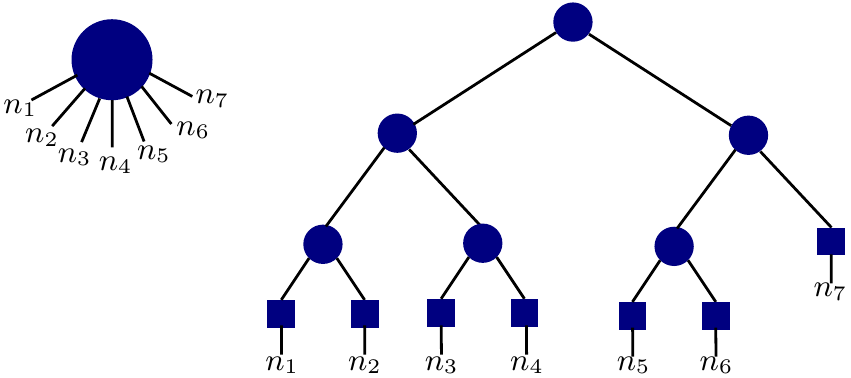}
    \caption{An illustration of the Hierarchical Tucker Network. An input \mbox{order-7} tensor (left) and its decomposition tree $HT$ (right). Each circle node of $HT$ (except the root) is  an order-3 ``core'' tensor, and each square is a matrix. An edge connecting two nodes denotes a contraction operation.  \label{fig:tree}}
\end{figure}

We assume we are given an order-$d$ input tensor $\cal T$ to be transformed into a BHTN ${\cal HT}$. A visualization of the structure of a BHTN is given in Figure~\ref{fig:tree}. We use the same notation ${\cal HT}$ both for the BHTN as well as its associated decomposition tree. We describe a recursive algorithm, and assume that the tensor at the current level of recursion is $T(n_1,\ldots,n_s,q)$, where $n_i$ is the size in the $i$-th dimension and $q$ is the dimension to connect $T$, through a contraction operation, to the higher-level factor in the decomposition tree. We assume that $q=1$ corresponds to  level 0 of the decomposition tree (i.e., we create a dummy dimension of the input tensor, for uniformity of representation),  and for all other levels we  have $q>1$. Let $i$ denote the current iteration level. The algorithm consists of a sequence of reshaping and splitting operations, and the output is a subtree $HT$ of ${\cal HT}$ corresponding to a factorization of $T$.

Initially, we reshape $T$ into a matrix $M=M(n_1,\ldots,n_{s/2},n_{s/2+1},\ldots,n_s q)$. Assuming $s > 3$, the following recursion is admissible.

First, we split $M$ into two matrices of specified dimensions, using a Boolean matrix factorization algorithm, leading to
\begin{align}
    M \rightarrow M'(n_1,\ldots,n_{s/2},r_{(1,s/2)}) * M''(r_{(1,s/2)},n_{s/2+1},\ldots,n_s q).
    \label{eq:step1}
\end{align}
We will use $M'$ to compute (by recursion) the left branch (left subsubtree) of the decomposition subtree $HT$ corresponding to $T$, while $M''$ will be used to define the right branch of $HT$, as well as one order-3 tensor, called \textit{core}, which will be the root of $HT$ and which connects those two (left and right) branches, and also connects $HT$ to its parent 3-d tensor in ${\cal HT}$.

Second, we use reshaping to move the dimension $q$ from the columns to the rows of $M''$, resulting in
\begin{align}
    \eqref{eq:step1} \rightarrow M'(n_1,\ldots,n_{s/2},r_{(1,s/2)}) * M''(q r_{(1,s/2)},n_{s/2+1},\ldots,n_s).
    \label{eq:step2}
\end{align}
By extracting the core from $M''$, we leave $M'$ unchanged and are left with
\begin{align}
    \eqref{eq:step2} \rightarrow M'(n_1,\ldots,n_{s/2},r_{(1,s/2)}) &* M_\text{core}''(q r_{(1,s/2)},r_{(s/2+1,s)})\nonumber \\ 
    &* M_\text{right}''(r_{(s/2+1,s)},n_{s/2+1},\ldots,n_s).
    \label{eq:step3}
\end{align}
Recursively applying this decomposition to the left and right subtrees, and reshaping back into tensors, eventually yields
\begin{align}
    \label{eq:step4}
    \eqref{eq:step3} &\rightarrow HT_\text{left}([1,s/2],r_{(1,s/2)}) \times  T_\text{core}(q,r_{(1,s/2)},r_{(s/2+1,s)})\nonumber  \\
     &\hspace{4.1cm}\times HT_\text{right}([s/2+1,s],r_{(s/2+1,s)})\\
    &= HT([1,s],q), 
\end{align}
where we use $[k_1,k_2]$ to denote the sequence $k_1,k_1+1,\dots,k_2$.

The decomposition outlined here allows us to decompose a tensor which has been flattened out as a matrix so long as $s>3$. For $s \leq 3$, the decomposition is constructed explicitly.

As can be seen in the algorithm's description, there are two types of operations: reshaping, which can be implementing by reordering the elements of  the tensor/matrix in a straightforward manner, and factorization of a Boolean matrix into a product of two Boolean matrices, which is a more complex problem and will be considered in the following subsection.

\subsection{Iterative Boolean matrix factorization}
Suppose we are given a Boolean matrix $X \in \B^{n \times m}$, which we aim to represent  as a product, $X= AB$, where $A$ and $B$ are also Boolean matrices, and we will reduce the latter problem  into a series of problems suitable for solving on D-Wave.
We can approximately solve this problem by using an iterative minimization procedure,
\begin{align}
	A &= \arg\min_Y d(X,YB),\label{eq:iteration0}\\
	B &= \arg\min_Y d(X,AY),
    \label{eq:iteration}
\end{align}
where $d(\cdot,\cdot)$ denotes the Hamming distance. Note that when iteratively solving eq.~\eqref{eq:iteration0}-\eqref{eq:iteration} both minimizations can be performed with one subroutine by taking transposes, hence we focus only on solving eq.~\eqref{eq:iteration}.
This problem can be further reduced into  a single-column Boolean factorization problem, i.e.,
\begin{equation}
   B_i = \arg\min_{y} d(X_i,Ay) \mbox{ for $i \in \{1,\ldots,m\}$,} 
\end{equation}
for $y=Y_i$, since, for $A$ fixed, values of the $i$-th column of $X=AY$ depend only on values of the $i$-th column of $Y$.
Let $T_i = \{j: X_{ji}=1 \}$ be the set of all indices with entry \textit{true} in column $X_i$, and likewise let $F_i = \{j: X_{ji}=0 \}$ be the set of all indices with entry \textit{false} in column $X_i$.
Then we can write
\begin{equation}
    d(X_i,Ay) = C - \sum_{j \in T_i} f( (A^\top)_j \cdot y) + \sum_{j \in F_i} f( (A^\top)_j \cdot y),
    \label{eq:hubo}
\end{equation}
where $f(x_1,\ldots,x_n) = 1 - \prod_{i=1}^n (1-x_i)$ and $C$ is equal to the number of non-zero entries in column $X_i$. Moreover, note that $f( (A^\top)_j \cdot y)$ can be reformulated as $f(y_{j_1},y_{j_2},\ldots,y_{j_r'})$, where $\{j_1,j_2,\ldots,j_{r'}\} = \{j:A_{ji}=1\}$. This makes eq.~\eqref{eq:hubo} a so-called higher order binary optimization (HUBO) problem in a (possibly complete) subset of the variables $\{y_1,y_2,\ldots,y_r\}$, where $r$ is the rank of the factorization.

To solve eq.~\eqref{eq:hubo} on D-Wave, we first need to convert it into a quadratic unconstrained binary optimization (QUBO) problem. A QUBO in $n$ binary variables is given by a minimization of the following function,
\begin{align}
    \sum_{i=1}^n a_i x_i + \sum_{i<j} a_{ij}x_i x_j,
    \label{eq:qubo}
\end{align}
where $x_i \in \{0,1\}$ for $i \in \{1,\ldots,n\}$ are the variables, and $a_i, a_i$ $\in \R$, for $i,j \in \{1,\ldots,n\}$, are chosen by the user to define the problem under investigation. Note that while the objective function of a QUBO is a polynomial of degree two, the degree of a HUBO can be arbitrarily large, e.g., a HUBO can include terms such as $a_{ijk}x_ix_jx_k$.

Conversions from higher order problems to QUBO are part of the D-Wave API \cite{hubo1,hubo2}. The penalty factor (called "strength" in the D-Wave documentation) for the HUBO to QUBO conversion was always chosen as the maximum of the absolute value of any HUBO coefficient. In our experiments this choice preserved the minimum ground state solution while not causing the coefficients to become too large.

To solve the QUBO obtained after conversion, its coefficients are mapped onto the quantum chip, and a number of parameters such as the annealing time, the number of anneals, and other embedding-related ones are set by the user. Valid annealing times for D-Wave 2000Q are between 1 and 2000 microseconds, and the number of anneals specifies how many times the annealing process is repeated in a single annealer call. Since current quantum technology is noisy, and the outcome of quantum measuring is not deterministic, hundreds or thousands of anneals are usually done in a single call to the annealer, and the best of the proposed solutions is chosen.

\section{Experiments}
\label{sec:experiments}
This section presents our experimental results for our BHTN implementaion. All results are computed with the algorithms of Section~\ref{sec:methods}.

Since these algorithms call the D-Wave 2000Q, which allows for the specification of several tuning parameters, we first calibrate a variety of D-Wave parameters (in particular, annealing time, an embedding parameter called chain strength, and the number of anneals). The details of calibration results are omitted here for brevity of the exposition.

We are interested in investigating how our algorithm performs if we vary certain properties of the tensor being factored (characterized by the order and size of the tensor) and if the factorization model is different (i.e., if we vary the rank). In the following, we always fix two of these three parameters and vary the third. We measure the error rates and the QPU (Quantum Processing Unit) times.

Additionally, we perform each experiment twice, once without adding noise to each tensor, and once with noise. In the latter case, we randomly apply independent bit flips with a probability of $0.01$ to the tensor under investigation. We generate the factors of the BHTN of our test tensors as Boolean matrices with Bernoulli entries (entry $1$ chosen with probability $p$, and entry $0$ chosen with probability $1-p$), which are then reshaped as tensors. By multiplying them into the final input tensor using the BHTN format, we make sure that each of our test tensors (before adding noise) has at least one exact BHTN representation. For each new test problem, we use an individual probability $p$ which we sample uniformly from $\{0.1,0.2,\ldots,0.9\}$. Also, we ensure that each tensor we generate (either without added noise, or with added noise) does not solely contain only entries $0$ (false) or entries $1$ (true).

After having generated a Boolean test tensor $\cal T$, we apply the algorithm of Section~\ref{sec:methods} to it and record the BHTN. We then compare $\cal T$ and the tensor obtained by multiplying the tensors of the computed BHTN by counting the number of elements that are dissimilar between both tensors, and divide that count by the number of elements in $\cal T$.

\begin{figure}[t]
    \centering
    \includegraphics[width=0.45\textwidth]{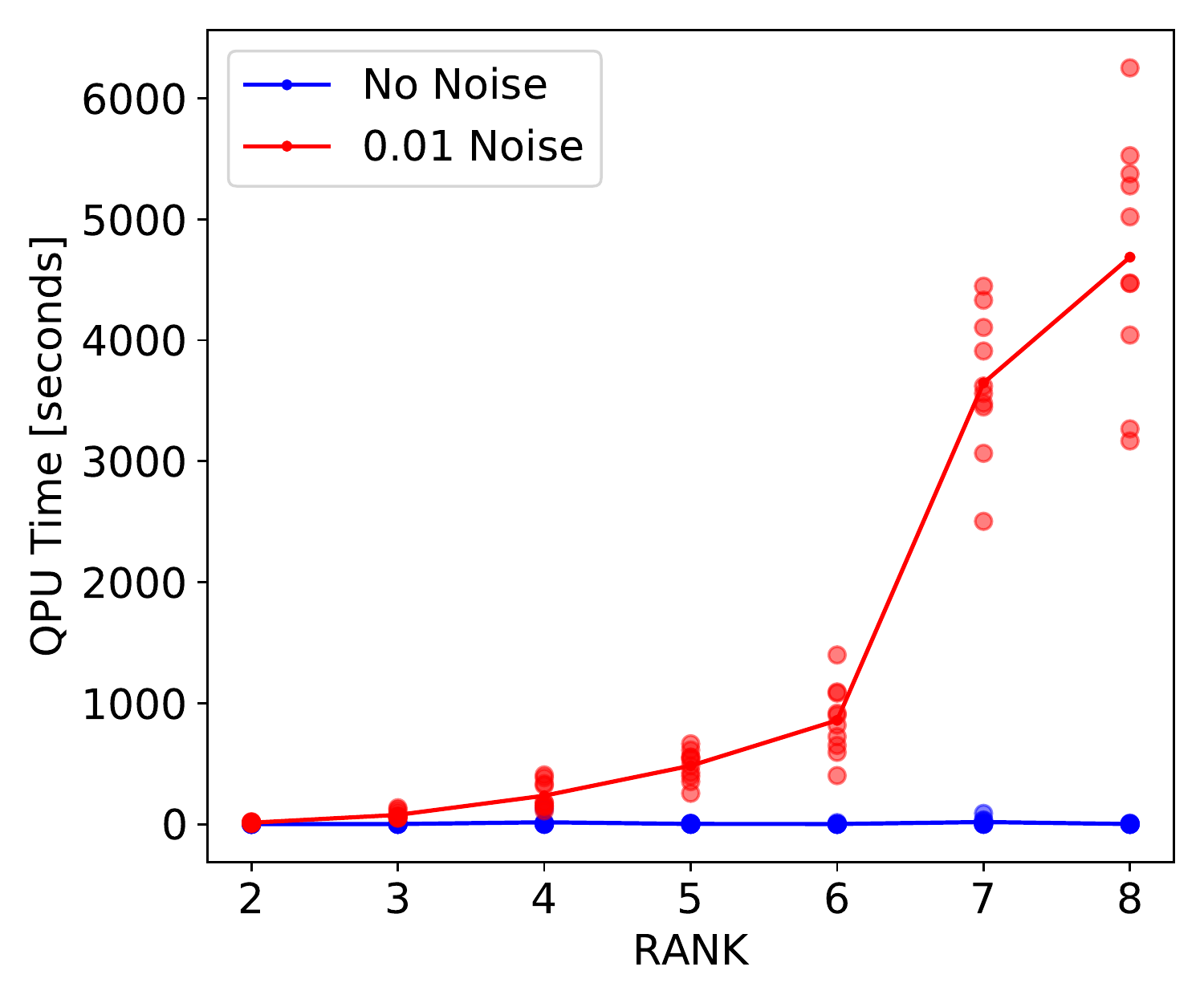}
    \includegraphics[width=0.45\textwidth]{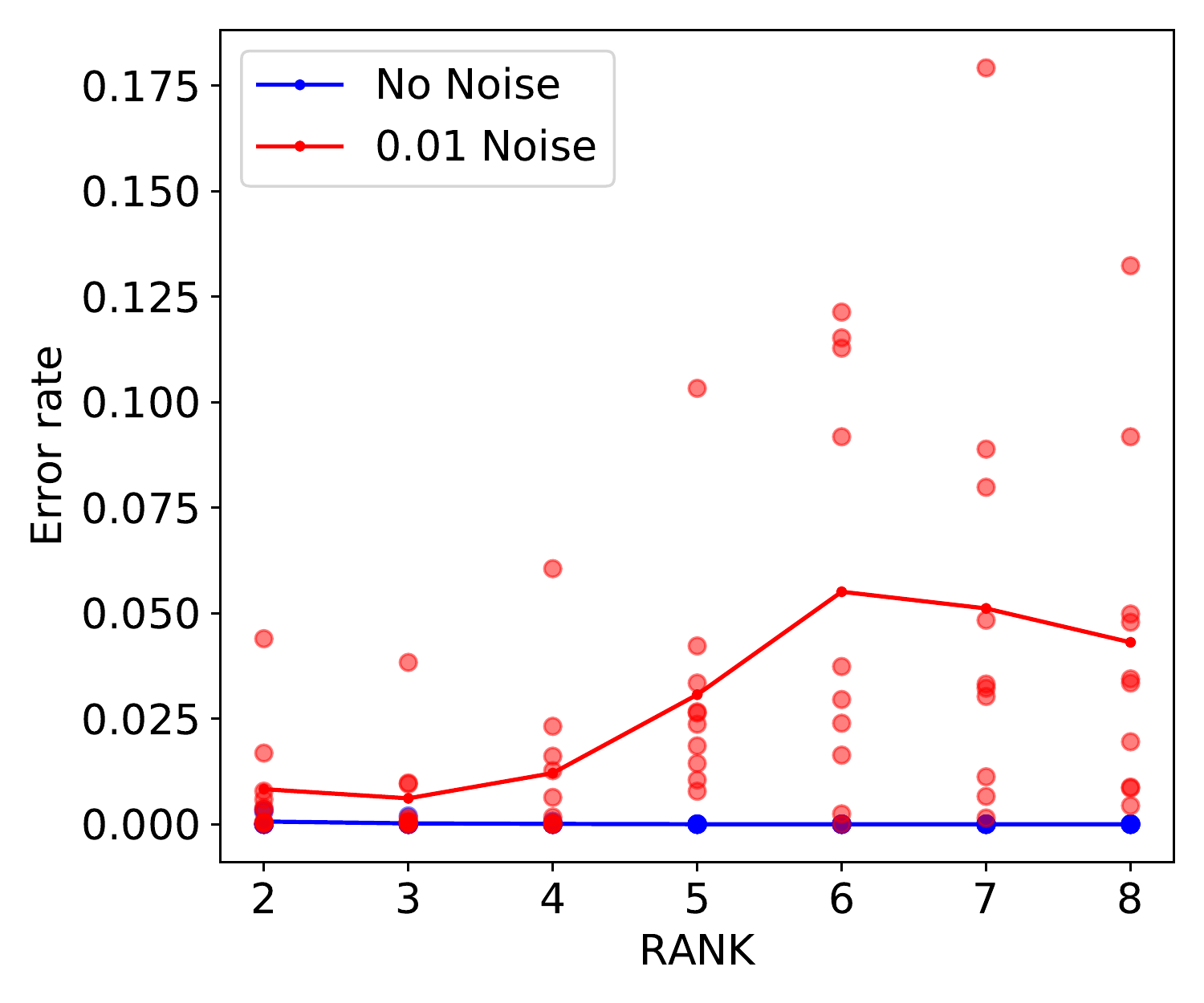}\\
    \caption{QPU times (left) and error rates (right) as a function of the rank while keeping the order $4$ and the size $8$ fixed. Tensor without added noise in blue, and tensor with added noise in red. Solid lines connect the means of the scatter plot values.}
    \label{fig:factor_tensor_vary_rank}
\end{figure}
Figure~\ref{fig:factor_tensor_vary_rank} shows the results as we vary the rank of the tensor while keeping the order, $4$, and the size, $8$, fixed. The QPU time and error rate both show little or no dependence on the rank for the scenario without added noise (blue curves). With noise, we observe that the QPU time and the error rates have a dependence on the rank. Note that the QPU time goes up with the rank due to the larger number of anneals needed to achieve the same error rate.

\begin{figure}[t]
    \centering
    \includegraphics[width=0.45\textwidth]{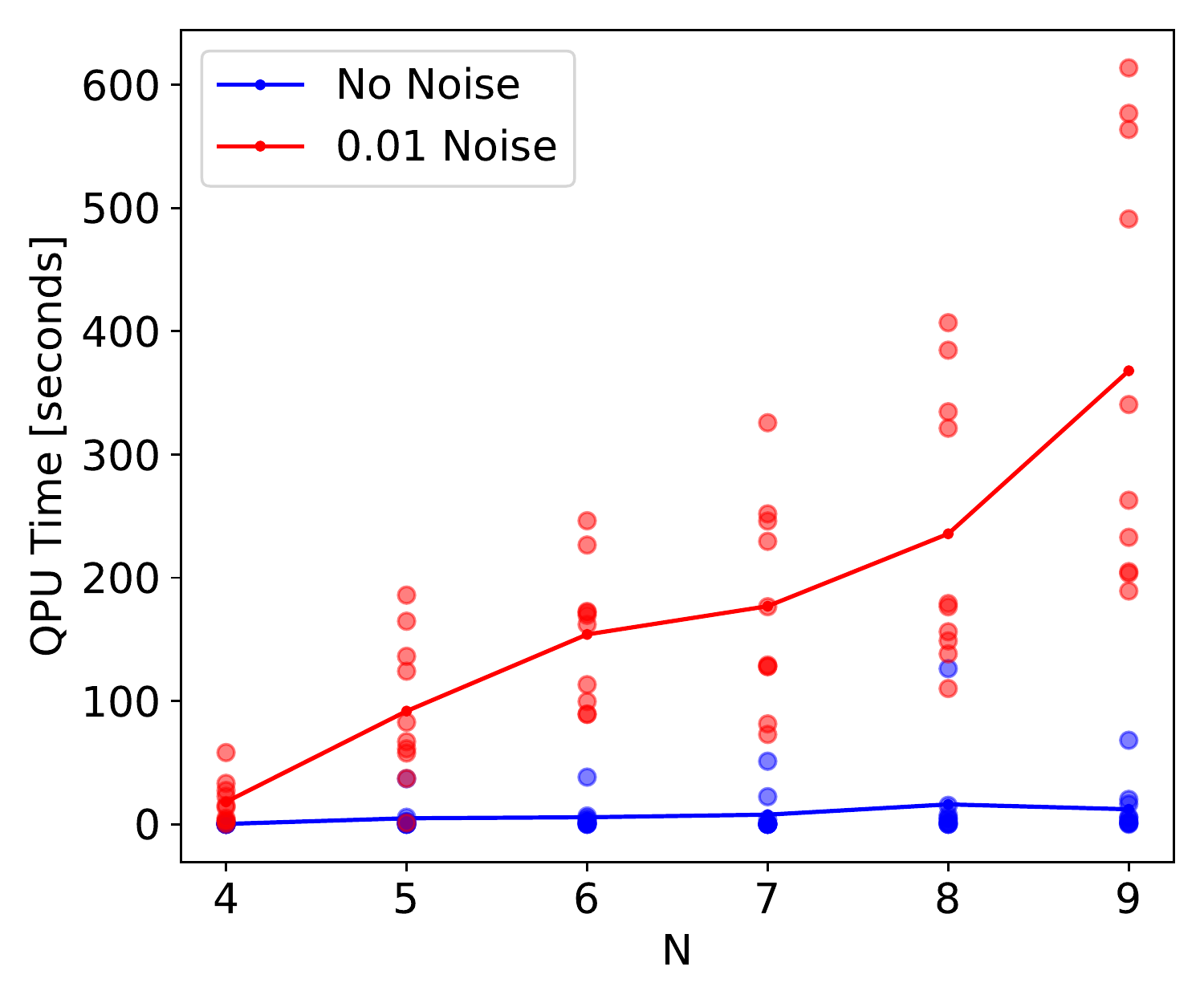}
    \includegraphics[width=0.45\textwidth]{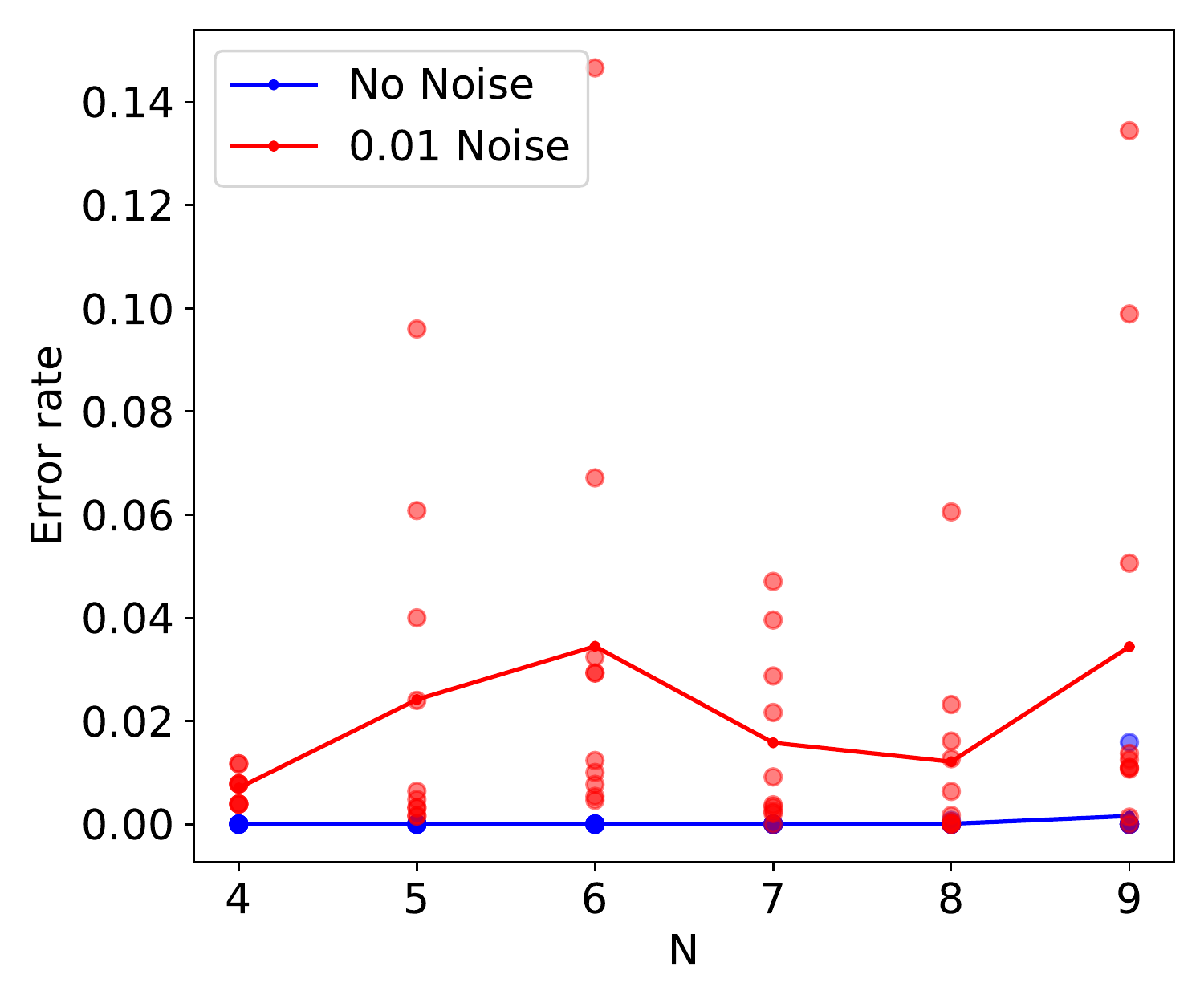}\\
    \caption{QPU times (left) and error rates (right) as a function of the size while keeping the rank $4$ and the order $4$ fixed. Tensor without added noise in blue, and tensor with added noise in red. Solid lines connect the means of the scatter plot values.}
    \label{fig:factor_tensor_vary_size}
\end{figure}
Figure~\ref{fig:factor_tensor_vary_size} shows the results for the scenario in which we fix both the rank and the order at $4$, and vary the size of the tensor. Without added noise, we observe very low QPU times and error rates. With added noise, we observe a roughly linear increase in QPU time as a function of the size. Error rates for the scenario with added noise fluctuate without significantly changing with $N$,
increasing from size 4 to 6, going down from  6 to 8, and up again at size 9.

\begin{figure}[t]
    \centering
    \includegraphics[width=0.45\textwidth]{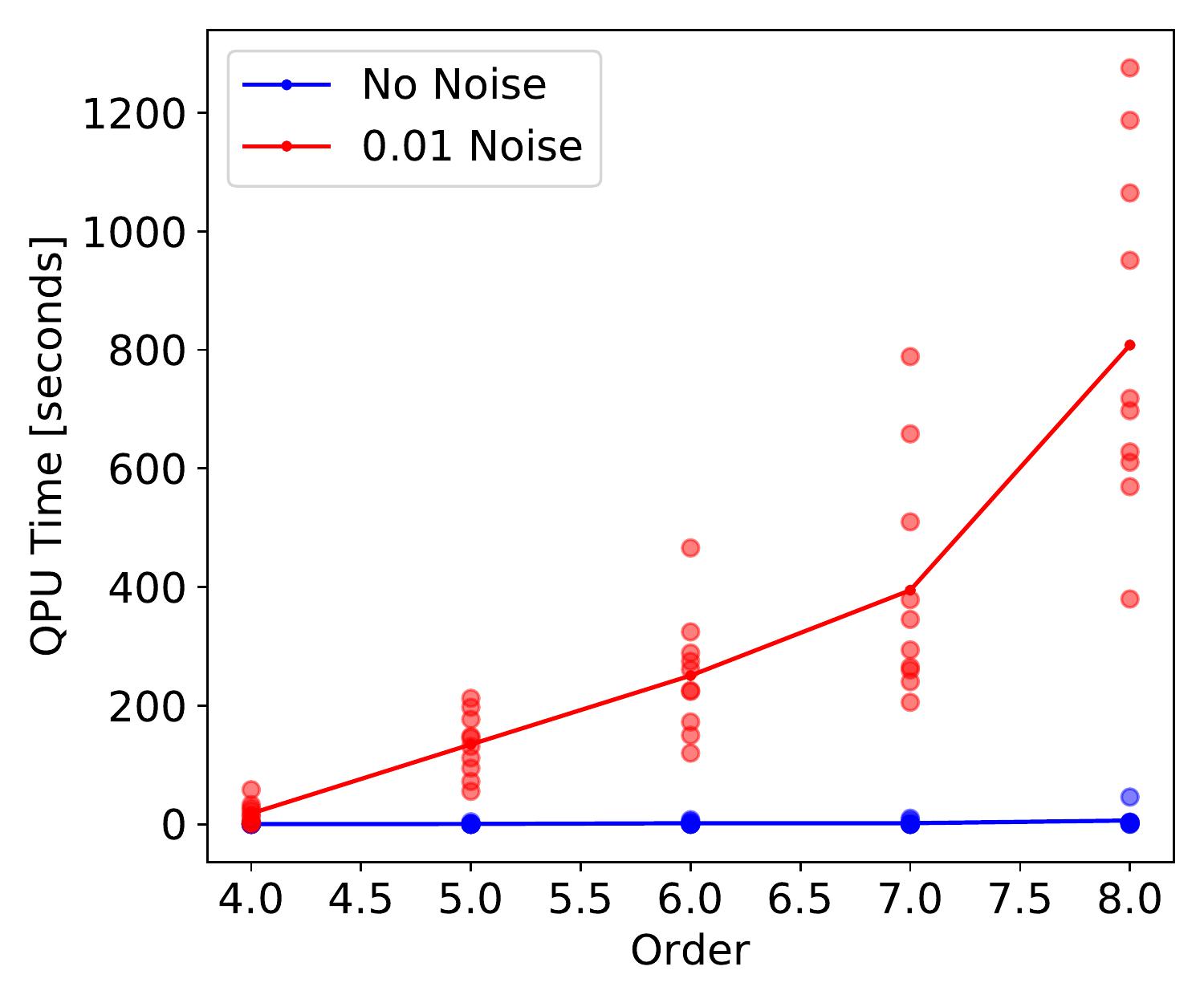}
    \includegraphics[width=0.45\textwidth]{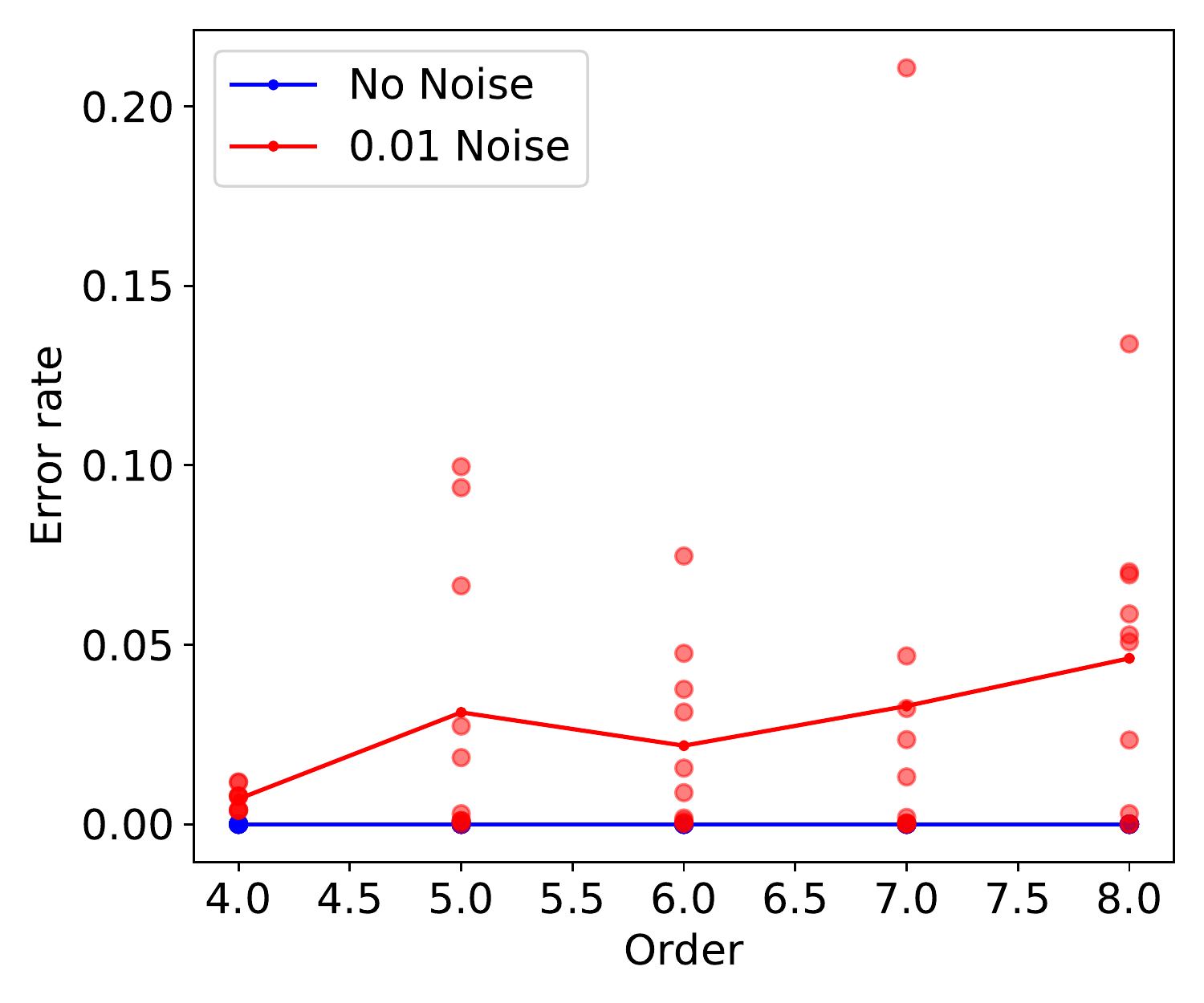}\\
    \caption{QPU times (left) and error rates (right) as a function of the order while keeping the rank $4$ and the size $4$ fixed. Tensor without added noise in blue, and tensor with added noise in red. Solid lines connect the means of the scatter plot values.}
    \label{fig:factor_tensor_vary_order}
\end{figure}
Finally, we look at the behavior of the algorithm of Section~\ref{sec:methods} as we vary the order of the tensor. We observe that without noise both the error rate and the QPU time remain fairly constant and close to zero. With noise, we observe a linearly increasing trend for the QPU time measurements. The error rate plot for tensors with noise has a less clear trend as a function of order, although we also observe an increase in error rates as the order increases.

\section{Discussion}
\label{sec:discussion}
This article considers computing Boolean Hierarchical Tucker Networks (BHTNs) on quantum annealers. To this end, we introduce an algorithm that allows us to break down an input tensor as a product of smaller-order Boolean tensors. The latter can again be factored in a recursive fashion to produce a BHTN.

On a lower implementation level, we consider a Boolean matrix factorization problem, for which we present an iterative factorization algorithm that reformulates the Boolean matrix factorization as the minimization of a (higher order) unconstrained binary optimization problem. This problem can be solved efficiently on the D-Wave 2000Q quantum annealer after converting it into a quadratic unconstrained binary optimization (QUBO) problem, the type of function the D-Wave 2000Q quantum annealer is designed to minimize.

An experimental section considers the factorization of synthetic Boolean tensors of varying ranks, sizes, and orders. We show that the D-Wave 2000Q annealer, in connection with our strategy, allows one to accurately compute a BHTN for an initial  tensor of up to order $8$.

\section*{Acknowledgments}
\label{sec:Acknowledgments}
The research presented in this article was supported by the Laboratory Directed Research and Development program of Los Alamos National Laboratory under the project numbers 20190020DR and 20190065DR.


\begin{thebibliography}{}
\bibitem[DeSantis et~al., 2020]{desantis2020factorizations}
DeSantis, D., Skau, E., and Alexandrov, B. (2020).
\newblock Factorizations of binary matrices--rank relations and the uniqueness
  of boolean decompositions.
\newblock {\em arXiv preprint arXiv:2012.10496}.

\bibitem[Golden and O'Malley, 2021]{golden2021reverse}
Golden, J. and O'Malley, D. (2021).
\newblock Reverse annealing for nonnegative/binary matrix factorization.
\newblock {\em Plos one}, 16(1):e0244026.

\bibitem[Inc., 2020a]{hubo2}
Inc., D.-W.~S. (2020a).
\newblock
  \href{https://docs.ocean.dwavesys.com/projects/dimod/en/latest/reference/generated/dimod.higherorder.utils.make_quadratic.html#dimod.higherorder.utils.make_quadratic}{Create
  a binary quadratic model from a higher order polynomial}.

\bibitem[Inc., 2020b]{hubo1}
Inc., D.-W.~S. (2020b).
\newblock
  \href{https://docs.ocean.dwavesys.com/projects/dimod/en/latest/reference/higherorder.html}{Higher-Order
  Models}.

\bibitem[Kolda and Bader, 2009]{kolda2009tensor}
Kolda, T.~G. and Bader, B.~W. (2009).
\newblock Tensor decompositions and applications.
\newblock {\em SIAM review}, 51(3):455--500.

\bibitem[Monson et~al., 1995]{Monson1995}
Monson, S.~D., Pullman, N.~J., and Rees, R. (1995).
\newblock {A Survey of Clique and Biclique Coverings and Factorizations of
  (0,1)-Matrices}.
\newblock {\em Bull. ICA}, 14:17--86.

\bibitem[O'Malley et~al., 2018]{o2018nonnegative}
O'Malley, D., Vesselinov, V., Alexandrov, B., and Alexandrov, L. (2018).
\newblock Nonnegative/binary matrix factorization with a d-wave quantum
  annealer.
\newblock {\em PloS one}, 13(12):e0206653.

\bibitem[Spearman, 1961]{spearman1961general}
Spearman, C. (1961).
\newblock ``{G}eneral intelligence,'' objectively determined and measured.
\newblock {\em The American Journal of Psychology}, 15.
\end{thebibliography}

\end{document}